\documentclass[prl,reprint,graphicx,amssymb,amsmath,superscriptaddress]{revtex4-1}

\usepackage{rotating}

\renewcommand{\section}[1]{\textit{#1}--}

\def \beq {\begin{equation}}
\def \eeq {\end{equation}}
\def \ba {\begin{eqnarray}}
\def \ea {\end{eqnarray}}

\begin{document}

\title{Quantum interface between an electrical circuit and a single atom}
\author{D. Kielpinski}
\affiliation{Centre for Quantum Dynamics, Griffith University, Nathan, QLD 4111, Australia}
\author{D. Kafri}
\affiliation{Joint Quantum Institute/NIST, College Park, MD, USA}
\author{M. J. Woolley}
\affiliation{Centre for Engineered Quantum Systems, School of Mathematics and Physics, The University of Queensland, St Lucia, QLD 4072, Australia}
\author{G. J. Milburn}
\affiliation{Centre for Engineered Quantum Systems, School of Mathematics and Physics, The University of Queensland, St Lucia, Australia 4072}
\author{J. M. Taylor}
\affiliation{Joint Quantum Institute/NIST, College Park, MD, USA}

\begin{abstract}
We show how to bridge the divide between atomic systems and electronic devices by engineering a coupling between the motion of a single ion and the quantized electric field of a resonant circuit. Our method can be used to couple the internal state of an ion to the quantized circuit with the same speed as the internal-state coupling between two ions. All the well-known quantum information protocols linking ion internal and motional states can be converted to protocols between circuit photons and ion internal states. Our results enable quantum interfaces between solid state qubits, atomic qubits, and light, and lay the groundwork for a direct quantum connection between electrical and atomic metrology standards.
\end{abstract}

\maketitle

Atomic systems are remarkably well suited to storage and processing of quantum information \cite{Monroe-atom-photon-QC-rev, Blatt-Wineland-ion-QIP-2008-rev}. However, their properties are tightly constrained by nature, causing difficulties in interfacing to other optical or electronic devices.  On the other hand, quantum electronic circuits, such as superconducting interference devices, may be easily engineered to the designer's specifications and are readily integrated with existing microelectronics \cite{Clarke-Wilhelm-SC-qubit-rev}. The naturally existing couplings between a single atom and a single microwave photon in a superconducting circuit are too weak for practical coherent interfaces. The coupling has been estimated at tens of Hz \cite{Verdu-Schmiedmayer-ultracold-atoms-SC-cavity}, much smaller than the decoherence rate of $10^3 \:\mbox{s}^{-1}$. For trapped ions, the coupling between the electric dipole induced by ion motion and the electric field of the superconducting circuit can be much larger, on the order of several hundred kHz. Unfortunately, this coupling is far off resonance. Motional frequencies of trapped ions are limited to tens of MHz, while any superconducting circuit must maintain GHz operating frequencies to avoid thermal noise, even in the extreme cryogenic environment of a dilution refrigerator.

\begin{figure}
\includegraphics[width=3.1in,bb=0 0 595 264]{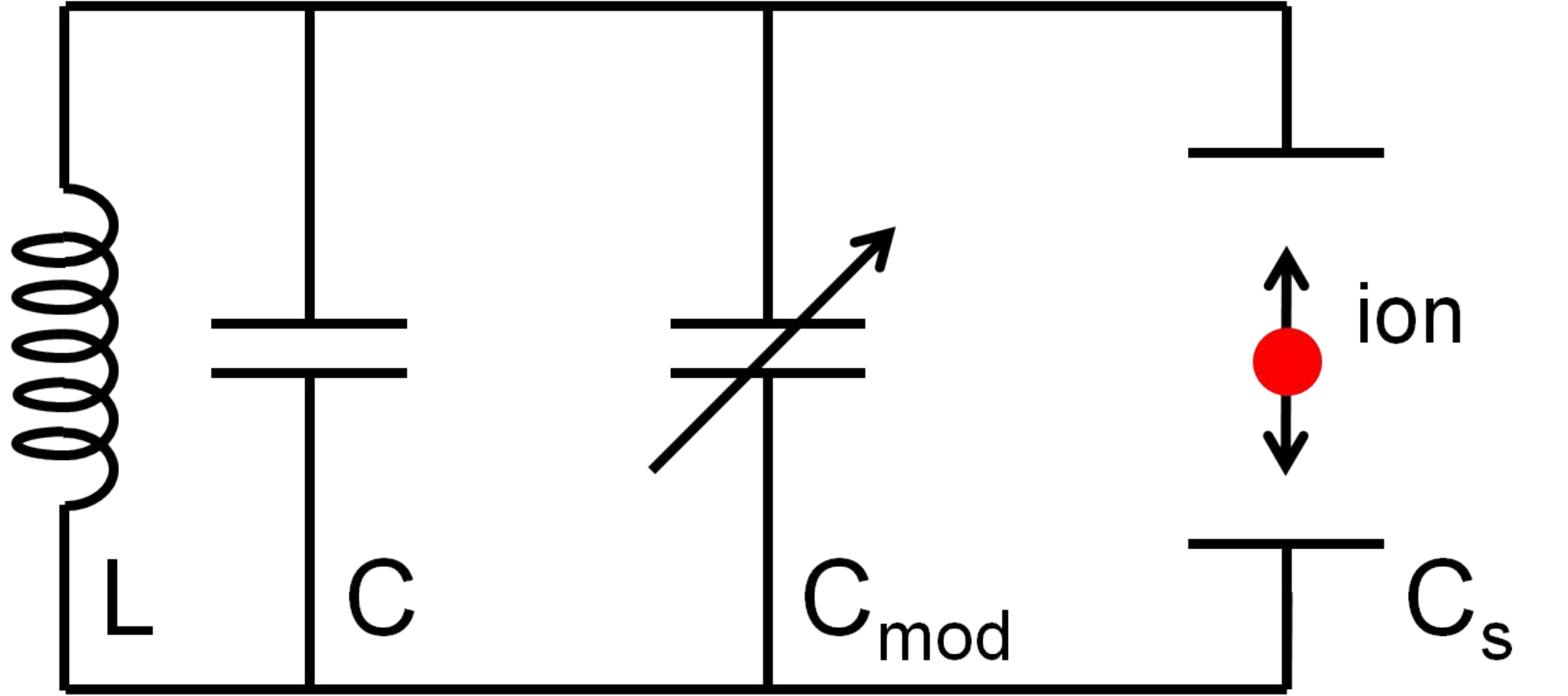}
\caption{Equivalent-circuit model of our scheme for ion-circuit coupling.}
\label{simpleschem}
\end{figure}

In this Letter, we propose a method to couple single trapped ions with microwave circuits, bridging the gap between the very different frequencies of ion motion and microwave photon by parametric modulation of the microwave frequency. The resulting coupling strength of $\sim 2 \pi \times 60$ kHz is sufficient for high-fidelity coherent operations and similar to the strength of currently obtained ion-ion couplings  \cite{Leibfried-Wineland-geometric-gate, Benhelm-Blatt-low-gate-error}. A simple model system illustrating the key concepts is shown in Fig. \ref{simpleschem}. Microwave photons reside in a superconducting LC circuit with natural frequency $\omega_\mathrm{LC} = 1/\sqrt{LC} \approx 1$ GHz. A single ion is confined within the capacitor $C_s$ and can oscillate at the motional frequency $\omega_i \approx 10$ MHz. The circuit voltage across $C_s$ generates an electric field that couples to the ion's motional electric dipole. Modulating the circuit capacitance by $C_\mathrm{mod}$ at a frequency $\nu$ causes the superconducting voltage to acquire sidebands at frequencies $\omega_\mathrm{LC} \pm \nu$. The coupling between the superconducting circuit and the ion motion becomes resonant when $\omega_i \approx \omega_\mathrm{LC} - \nu$. The interaction Hamiltonian is then
\beq
H_\mathrm{int} = \hbar g \, a b^\dagger + \mbox{h.c.} \label{ionlc}
\eeq
where $a$ and $b$ are the annihilation operators of the microwave photon mode and the ion motional mode, respectively. As shown below, $g \sim 2 \pi \times 60$ kHz.

The coupling between the LC circuit and the ion motion allows us to generalize all the well-known protocols operating on ion spin and motion to protocols operating on ion spin and LC state. Ion spin-motion protocols based on laser \cite{Leibfried-Wineland-single-ion-quantum-rev} or microwave fields \cite{Ospelkaus-Wineland-microwave-quantum-logic, Timoney-Wunderlich-microwave-quantum-logic} now allow for generation of nearly arbitrary spin/motion entangled states. If the capacitance modulation is switched on for a time $T = \pi/(2 g)$, Eq. (\ref{ionlc}) shows that the mode operators evolve as $a(T) = -i b(0), b(T) = -i a(0)$, i.e., a perfect swap between LC and motional modes. For a unitary operator $U(b,\vec{\sigma})$ describing a protocol between the ion motion and the ion spin operator $\vec{\sigma}$, the sequence 1) swap LC/motion, 2) apply $U(b, \vec{\sigma})$, 3) swap LC/motion implements the same unitary $U(a,\vec{\sigma})$ between the LC and spin modes. By this means, one can establish a quantum communications channel between LC circuits in separate dewars, couple ion spins through a common LC circuit for large-scale quantum computing on a single chip, and perform Heisenberg-limited voltage metrology in the microwave domain by generating large Schr\"odinger cat states of the LC mode.

\begin{figure}
\includegraphics[width=\columnwidth,bb=0 0 558 410]{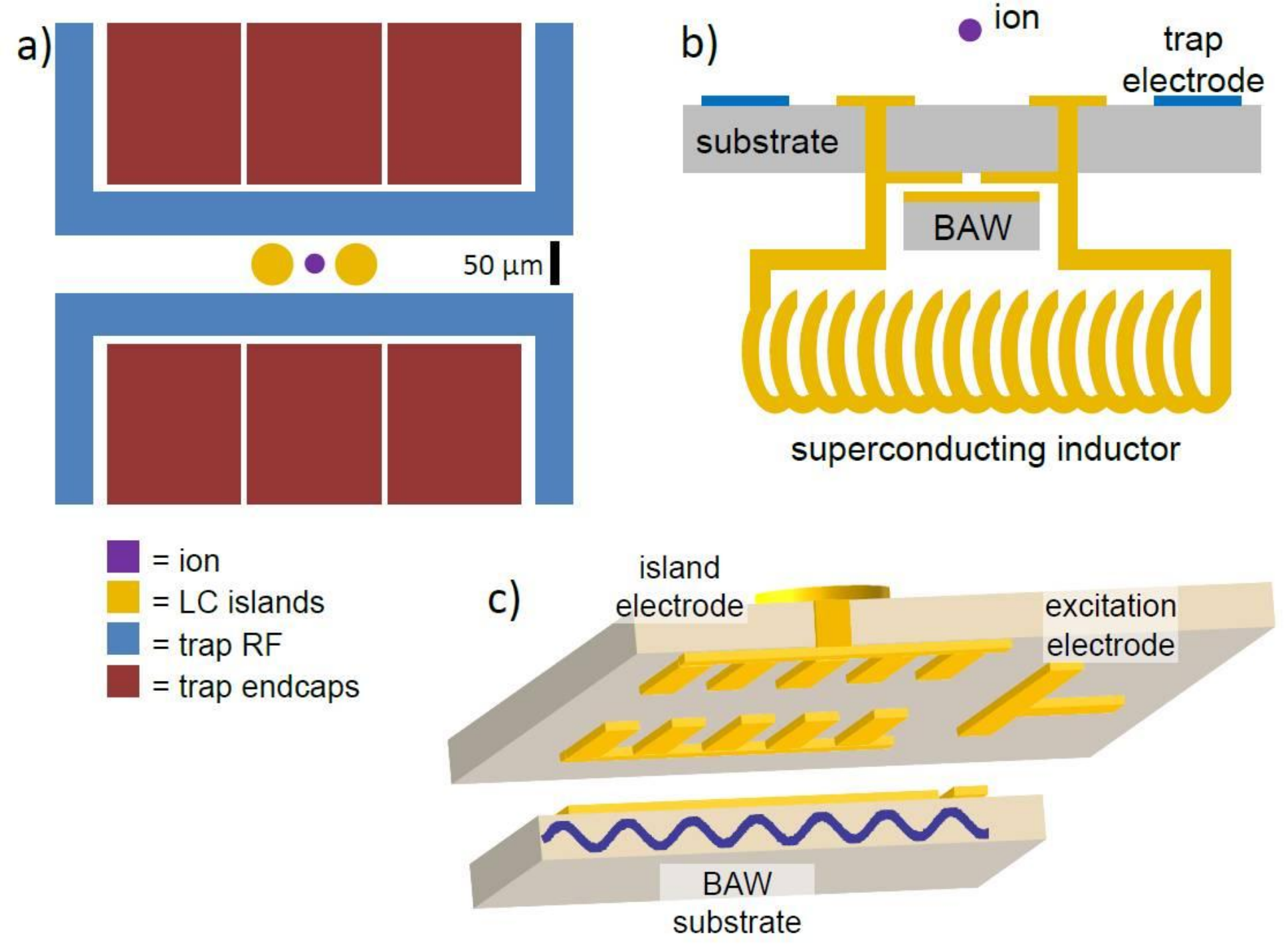}
\caption{Schematic of a device for coupling trapped ions to a microwave resonant circuit. a) Top view of surface ion trap showing RF and DC trapping electrodes. The ``LC island'' electrodes couple the ion motion to the LC circuit excitation. b) Side view of device, showing ion trap, superconducting inductor, and BAW device. c) Exploded side view of BAW device. Purple line: transverse displacement of BAW substrate due to classical driving.}
\label{schem}
\end{figure}

\textit{Realisation of ion-CQED coupling.--} Figure \ref{schem} shows a schematic of a device implementing the simple model described above. The device combines an ion trap, a microwave LC circuit, and a bulk-acoustic-wave (BAW) microelectromechanical modulator to coherently couple the quantized motion of trapped ions at MHz frequencies with microwave photons at 1 GHz.  We now describe a specific design to give real-world parameters relevant to our system.

Ions are confined above a planar electrode structure of a type now widely used for microfabricated trap arrays \cite{Chiaverini-Wineland-surface-trap-design}. Applying appropriate voltages to the electrodes generates RF electric fields, which provide a ponderomotive confining potential transverse to the trap axis, and DC fields that give rise to a harmonic potential along the axis. The ion trap parameters are taken to be typical for planar traps \cite{Seidelin-Wineland-surface-trap}, with ions confined at height $h = 25 \: \mu$m above the plane with an axial frequency $\omega_i$ of $2\pi \times 1$ MHz. For the commonly used $^9\mbox{Be}^+$ ion, the harmonic oscillator length is then $z_0 = \sqrt{\hbar/(2 m \omega_i)} = 24$ nm.

A superconducting inductor is attached to the island electrodes and a silicon bulk-acoustic-wave resonator (BAW) is mounted near the inductor (see the Supplemental Material for technical details). The inductance of 440 nH, combined with the total static circuit capacitance of $C_0 = 46$ fF, then yields $\omega_\mathrm{LC} = 1$ GHz, with characteristic impedance of $Z = 2.7 \: \mathrm{k}\Omega$. The zero-point charge fluctuation on the resonator is $q_0 = \sqrt{\hbar/(2 Z)} = 0.9$ electrons.

The ion-circuit coupling is provided through two coplanar islands near the ion position that are each connected to a terminal of the superconducting inductor. The microwave electric field between these islands couples to the ion motion along the trap axis through the electric dipole of the moving ion charge. To activate the ion-circuit coupling, one excites acoustic waves in the BAW at frequency $\nu_B \approx \omega_\mathrm{LC} - \omega_i$ by voltage driving of metallic electrodes on the BAW surface. The modulation of the BAW-substrate gap distance provides the desired capacitance modulation.

The classical dipole interaction energy of the ion due to the axial electric field $E_z$ from the island electrodes is
\begin{equation}
U_\mathrm{cl} = e z E_z = \frac{e \zeta}{h} z \: V = \frac{e \zeta}{h C} z Q
\end{equation}
where $h$ is the ion height, $V$ is the voltage between the islands, $\zeta$ is a dimensionless constant of order unity set by the electrode geometry, $C$ is the total circuit capacitance, and $Q$ is the total charge on the circuit. Simulation of the electric field near the island electrodes gives $\zeta = 0.25$. The BAW drive modulates the capacitance as $C = C_0 (1 + \eta \sin \nu t)$ with modulation depth $\eta = 0.3$, so that
\beq
U_\mathrm{cl}(Q,z,t) = \frac{e \zeta}{h C} (1 - \eta \sin \nu t) z Q
\label{classpot}
\eeq

We now quantize the LC and ion motion, but keep the BAW motion classical. In the rotating frame with respect to LC and motion, the total Hamiltonian of the ion-LC system becomes (for details of the calculation, see the Supplemental Material)
\beq
H_\mathrm{int}/\hbar = \frac{2 i g_0 \eta}{3} e^{-i \Delta t} a b^\dagger + \mbox{h.c.} \label{motlc}
\eeq
where $g_0 = e \zeta z_0 q_0 / (h C_0)$ and $\Delta \equiv \nu - (\omega_\mathrm{LC} - \omega_i)$. For the numerical parameters given above, $g_0 = 2 \pi \times 200$ kHz and $\eta = 0.3$, giving $g = 2 \pi \times 60$ kHz.

Because the BAW is only used as a parametric drive in our scheme, it contributes negligible noise to the LC and motional modes. To first order, the only semiclassical effect of the BAW is variation in the coupling parameter $\eta$ between the ion motion and the LC--there is no direct (linear) coupling between the motion of the BAW and these other two variables. Hence parametric heating is the main source of noise added by the BAW. While thermal motion of the BAW can in principle produce parametric heating, in most practical settings errors in $\eta$ will be determined by classical control errors in setting the BAW amplitude.

The chief quantum noise contribution of the BAW arises from the entanglement induced by the LC and ion systems with the BAW and, indirectly, its environment.  This entanglement occurs via the parametric coupling, and manifests as a static displacement of the BAW that depends on LC photon number $n_{LC}$. The displacement can be estimated as
\beq
\zeta_\mathrm{LC} \sim \frac{x_B}{\zeta_0} \frac{n_\mathrm{LC} \omega_\mathrm{LC}}{\omega_\mathrm{LC} - \nu + i \kappa_B/2}
\eeq
where $x_B$ is the harmonic oscillator length of the BAW and $\kappa_B$ is the BAW damping rate. For typical BAW parameters at $\nu \sim 1$ GHz, one finds $x_B \sim 10^{-16}$ m and $\kappa_B \sim 100$ kHz, so that $\zeta_\mathrm{LC}/\zeta_0 \lesssim 10^{-3}$ even for $n_\mathrm{LC} \sim 100$. Hence the BAW contributes negligible quantum noise for our purposes.

\textit{LC/spin protocols.--} LC/spin protocols can be executed by the swapping method with fidelity well over 95\%. $Q$ as high as $5 \times 10^5$ have been reported for an LC circuit \cite{Kim-Wellstood-Rb-hyperfine-SC-resonator}, giving a decoherence rate of 2 $\mbox{ms}^{-1}$. Motional decoherence rates of 0.5 s$^{-1}$ have been demonstrated in a cryogenically cooled ion trap with an ion height of $150 \: \mu$m and 1 MHZ motional frequency \cite{Labaziewicz-Chuang-cryogenically-suppressed-heating}. This rate scales as $\sim 1/d^4$ \cite{Turchette-Wineland-ion-heating}, so at our $25 \:\mu$m height, we estimate a rate of 0.5 $\mbox{ms}^{-1}$. Spin decoherence is negligible on these timescales \cite{Leibfried-Wineland-single-ion-quantum-rev}. Hence the overall decoherence rate is 2.5 $\mbox{ms}^{-1}$, limited by LC damping. The total spin/LC operation requires two LC/motion swaps, each taking 3 $\mu$s, and the spin/motion protocol, with typical Rabi frequency $\Omega_0 \sim 2 \pi \times 100$ kHz \cite{Leibfried-Wineland-single-ion-quantum-rev}. A typical spin/motion protocol requires approximately a $\pi/2$-pulse time, so the total time required for the LC/spin protocol is 10 $\mu$s. The infidelity is given by the ratio of decoherence rate to operation rate, i.e., 0.03.

Quantum interfaces between LC and spin can be achieved through a Jaynes-Cummings spin/motion interaction \cite{Leibfried-Wineland-single-ion-quantum-rev}. If the LC mode is regarded as the microwave analog of a linear-optical qubit, this interaction serves as a quantum logic interface between ion spin and single-rail microwave photon qubits. A $\pi/2$ pulse of the interaction performs an LC/spin CNOT gate.

An alternative LC-spin quantum interface swaps spin-dependent displacement of the ion motion into the LC mode. The unitary evolution for such a protocol is given by
\ba
U_\mathrm{eff}(\alpha) = \exp \left[ (\alpha a + \alpha^* a^\dagger) \sigma_x \right] \label{spindisp}
\ea
Such an operation could be used to teleport a superposition of spin states into a superposition of coherent LC states.

\begin{figure}
\includegraphics[width=\columnwidth,bb=0 0 715 265]{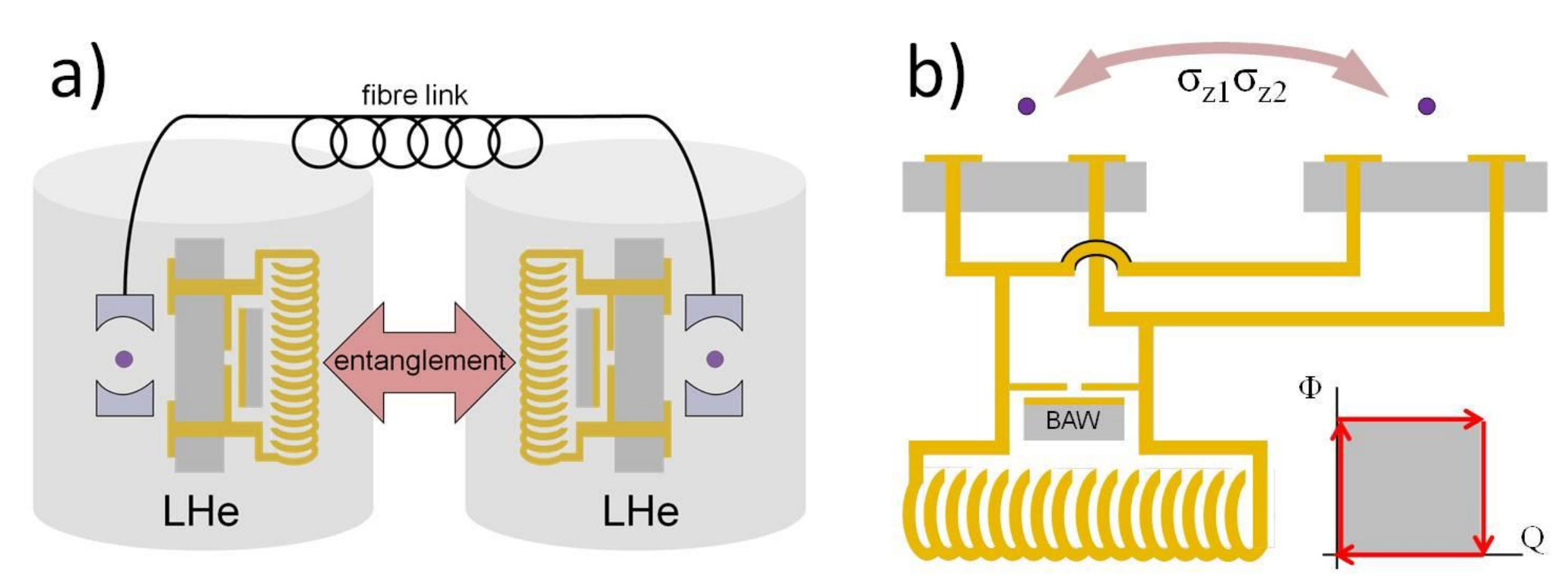}
\caption{Quantum buses enabled by LC/motion coupling. a) Quantum communication between LC circuits in independent cryogenic environments. b) Nonlocal ion spin-spin gates on a single chip.}
\label{busfig}
\end{figure}

These interactions allow us to use the LC and ion modes as quantum buses for more complex tasks. The Jaynes-Cummings LC/spin interaction enables quantum communication between LC circuits in independent cryogenic environments, as shown in Figure \ref{busfig}(a). Each ion interface is controllably coupled to a high-finesse optical resonator. After LC/spin coupling, the spin is mapped to the polarisation state of an outgoing optical photon, as in recent experiments \cite{Keller-Walther-ion-cavity-single-photon}. Overall, this set of operations coherently couples the microwave photon state to the optical domain. In particular, one could entangle independent superconducting qubits via conditional photon measurements \cite{Olmschenk-Monroe-separated-ion-teleportation}. This task is impossible using direct microwave signaling, owing to the thermal noise of microwave links at room temperature. \\

The spin-dependent LC displacement lets perform nonlocal ion spin-spin gates on a single chip through a shared LC mode (Fig \ref{busfig}(b)). The spin-spin bus is based on the spin-dependent displacement operation $D(\alpha q \sigma_y)$. Two ions (with identical phonon frequencies) in multiple traps are capacitively coupled to the LC circuit. Performing four spin-dependent displacements that enclose a square in phase space with side length $L = \alpha J_z $, one picks up a phase of $2 |L|^2 = 4  |\alpha|^2 (1 + \sigma_z^{(1)} \sigma_z^{(2)})$, which is the desired spin-spin interaction \cite{Milburn-James-phase-gate}. Along with single qubit gates, this interaction provides a sufficient gate set for universal quantum computation \cite{Barenco-Weinfurter-universal-gate-set}. \\

Spin-dependent displacement of the LC causes a superposition of spin states to evolve into a superposition of LC coherent states. Such superposed coherent states can detect field displacements with Heisenberg-limited sensitivity \cite{Munro-Braunstein-schrodinger-cat-heisenberg-limit}. In the present context, these states enable Heisenberg-limited metrology of small voltages at microwave frequencies. The mean photon number in the generated state can exceed $100$ for our parameters. The RMS voltage is then $\sim 0.1$ mV and can be estimated at sub-$\mu$V precision in a single shot.

\textit{Resistance to motional heating.--} Rapid technical advances in superconducting circuits mean that the LC coherence time may substantially increase in the near future, leaving ion motional heating as the primary source of decoherence. We have developed a modified LC/spin coupling scheme similar to \cite{Molmer-Sorensen-multiparticle-entangling-gate, Milburn-James-phase-gate} that resists ion heating. We simultaneously apply bichromatic LC/motion and motion/spin couplings at detunings $\pm \delta$ from the blue-sideband and red-sideband resonances. In the frame rotating with ion motion, the Hamiltonian becomes
\ba
H_\mathrm{int}/\hbar &=& \sqrt{2} M \left( x \cos \delta t + p \sin \delta t) \right) \label{msham} \\
M &\equiv& \frac{2 i \eta g_0}{3} q + \frac{\Omega_0}{4} \sigma_x
\ea
where $x = (b + b^\dagger)/\sqrt{2}$, $p = -i (b - b^\dagger)/\sqrt{2}$ are dimensionless motional operators, $q = (a + a^\dagger)/\sqrt{2}$ is the dimensionless charge operator, and we approximate $\Omega_0 \ll \delta \ll \omega_i$. The Hamiltonian (\ref{msham}) is identical to that of \cite{Sorensen-Molmer-bichromatic-gate-PRA}, except that the collective spin operator $J_y$ is replaced by the collective spin-LC operator $M$. The ion motional state undergoes $M$-dependent phase-space displacement along a closed trajectory, giving rise to an $M^2$ dependent geometric phase. At times $t_n = 2 \pi n / |\delta|$ with $n$ an integer, the evolution operator becomes simply
\beq
U_n = \exp \left[ -i \:\mbox{sign}(\delta) \frac{2 \pi n}{\delta^2} M^2 \right] \label{ums}
\eeq
The undesired $q^2$ term in Eq. (\ref{ums}) can be removed by a spin echo sequence $Z U_n^\dagger Z U_n$, where $Z = e^{-i \pi \sigma_z}$ is a fast $\pi$-pulse of the spin, and $U^\dagger$ is obtained by changing $\delta \rightarrow -\delta$. The overall time evolution is then $\exp(\alpha q \sigma_x)$ with $\alpha = -4 i \pi n g_0 \Omega_0 \eta / (3 \delta^2)$.

A straightforward modification of the arguments of S{\o}rensen and M{\o}lmer \cite{Sorensen-Molmer-bichromatic-gate-PRA} shows that this coupling can be made arbitrarily resistant to ion motional heating. The loss of fidelity due to heating is $\propto 1/\delta^2$ in the limit of low infidelity, while the effective coupling constant is $\propto 1/\delta$. Even if the heating rate is larger than the coupling constant, one can still achieve near-perfect coupling.

\textit{Outlook.--} Our ion-circuit coupling enables a powerful hybrid quantum system with operation speeds similar to those for ion spins. This system can perform nonlocal quantum gates between ions on a single chip, nonlocal quantum communication between electrical circuits, and Heisenberg-limited voltage metrology. The coupling can be made resistant to ion motional heating. Current experiments in classical ion-circuit coupling \cite{Ospelkaus-Wineland-microwave-quantum-logic, Timoney-Wunderlich-microwave-quantum-logic} can be extended in a natural way to realise our scheme.

This work was supported by the Australian Research Council under DP0773354 (Kielpinski), FF0458313 (Wiseman), the Centre of Excellence for Engineered Quantum Systems and Federation Fellowship funding (Milburn), and by the U.S. Army Research Office MURI award W911NF0910406 and DARPA QUASAR (Taylor). We acknowledge helpful conversations with Timothy Duty.


\begin{thebibliography}{21}%
\makeatletter
\providecommand \@ifxundefined [1]{%
 \@ifx{#1\undefined}
}%
\providecommand \@ifnum [1]{%
 \ifnum #1\expandafter \@firstoftwo
 \else \expandafter \@secondoftwo
 \fi
}%
\providecommand \@ifx [1]{%
 \ifx #1\expandafter \@firstoftwo
 \else \expandafter \@secondoftwo
 \fi
}%
\providecommand \natexlab [1]{#1}%
\providecommand \enquote  [1]{``#1''}%
\providecommand \bibnamefont  [1]{#1}%
\providecommand \bibfnamefont [1]{#1}%
\providecommand \citenamefont [1]{#1}%
\providecommand \href@noop [0]{\@secondoftwo}%
\providecommand \href [0]{\begingroup \@sanitize@url \@href}%
\providecommand \@href[1]{\@@startlink{#1}\@@href}%
\providecommand \@@href[1]{\endgroup#1\@@endlink}%
\providecommand \@sanitize@url [0]{\catcode `\\12\catcode `\$12\catcode
  `\&12\catcode `\#12\catcode `\^12\catcode `\_12\catcode `\%12\relax}%
\providecommand \@@startlink[1]{}%
\providecommand \@@endlink[0]{}%
\providecommand \url  [0]{\begingroup\@sanitize@url \@url }%
\providecommand \@url [1]{\endgroup\@href {#1}{\urlprefix }}%
\providecommand \urlprefix  [0]{URL }%
\providecommand \Eprint [0]{\href }%
\providecommand \doibase [0]{http://dx.doi.org/}%
\providecommand \selectlanguage [0]{\@gobble}%
\providecommand \bibinfo  [0]{\@secondoftwo}%
\providecommand \bibfield  [0]{\@secondoftwo}%
\providecommand \translation [1]{[#1]}%
\providecommand \BibitemOpen [0]{}%
\providecommand \bibitemStop [0]{}%
\providecommand \bibitemNoStop [0]{.\EOS\space}%
\providecommand \EOS [0]{\spacefactor3000\relax}%
\providecommand \BibitemShut  [1]{\csname bibitem#1\endcsname}%
\let\auto@bib@innerbib\@empty
\bibitem [{\citenamefont {Monroe}(2002)}]{Monroe-atom-photon-QC-rev}%
  \BibitemOpen
  \bibfield  {author} {\bibinfo {author} {\bibfnamefont {C.}~\bibnamefont
  {Monroe}},\ }\href@noop {} {\bibfield  {journal} {\bibinfo  {journal}
  {Nature}\ }\textbf {\bibinfo {volume} {416}},\ \bibinfo {pages} {238}
  (\bibinfo {year} {2002})}\BibitemShut {NoStop}%
\bibitem [{\citenamefont {Blatt}\ and\ \citenamefont
  {Wineland}(2008)}]{Blatt-Wineland-ion-QIP-2008-rev}%
  \BibitemOpen
  \bibfield  {author} {\bibinfo {author} {\bibfnamefont {R.}~\bibnamefont
  {Blatt}}\ and\ \bibinfo {author} {\bibfnamefont {D.}~\bibnamefont
  {Wineland}},\ }\href@noop {} {\bibfield  {journal} {\bibinfo  {journal}
  {Nature}\ }\textbf {\bibinfo {volume} {453}},\ \bibinfo {pages} {1008}
  (\bibinfo {year} {2008})}\BibitemShut {NoStop}%
\bibitem [{\citenamefont {Clarke}\ and\ \citenamefont
  {Wilhelm}(2008)}]{Clarke-Wilhelm-SC-qubit-rev}%
  \BibitemOpen
  \bibfield  {author} {\bibinfo {author} {\bibfnamefont {J.}~\bibnamefont
  {Clarke}}\ and\ \bibinfo {author} {\bibfnamefont {F.~K.}\ \bibnamefont
  {Wilhelm}},\ }\href@noop {} {\bibfield  {journal} {\bibinfo  {journal}
  {Nature}\ }\textbf {\bibinfo {volume} {453}},\ \bibinfo {pages} {1031}
  (\bibinfo {year} {2008})}\BibitemShut {NoStop}%
\bibitem [{\citenamefont {Verd{\'u}}\ \emph {et~al.}(2009)\citenamefont
  {Verd{\'u}}, \citenamefont {Zoubi}, \citenamefont {Koller}, \citenamefont
  {Majer}, \citenamefont {Ritsch},\ and\ \citenamefont
  {Schmiedmayer}}]{Verdu-Schmiedmayer-ultracold-atoms-SC-cavity}%
  \BibitemOpen
  \bibfield  {author} {\bibinfo {author} {\bibfnamefont {J.}~\bibnamefont
  {Verd{\'u}}}, \bibinfo {author} {\bibfnamefont {H.}~\bibnamefont {Zoubi}},
  \bibinfo {author} {\bibfnamefont {C.}~\bibnamefont {Koller}}, \bibinfo
  {author} {\bibfnamefont {J.}~\bibnamefont {Majer}}, \bibinfo {author}
  {\bibfnamefont {H.}~\bibnamefont {Ritsch}}, \ and\ \bibinfo {author}
  {\bibfnamefont {J.}~\bibnamefont {Schmiedmayer}},\ }\href@noop {} {\bibfield
  {journal} {\bibinfo  {journal} {Phys. Rev. Lett.}\ }\textbf {\bibinfo
  {volume} {103}},\ \bibinfo {pages} {043603} (\bibinfo {year}
  {2009})}\BibitemShut {NoStop}%
\bibitem [{\citenamefont {Leibfried}\ \emph
  {et~al.}(2003{\natexlab{a}})\citenamefont {Leibfried}, \citenamefont
  {DeMarco}, \citenamefont {Meyer}, \citenamefont {Lucas}, \citenamefont
  {Barrett}, \citenamefont {Britton}, \citenamefont {Itano}, \citenamefont
  {Jelenkovi{\'c}}, \citenamefont {Langer}, \citenamefont {Rosenband},\ and\
  \citenamefont {Wineland}}]{Leibfried-Wineland-geometric-gate}%
  \BibitemOpen
  \bibfield  {author} {\bibinfo {author} {\bibfnamefont {D.}~\bibnamefont
  {Leibfried}}, \bibinfo {author} {\bibfnamefont {B.}~\bibnamefont {DeMarco}},
  \bibinfo {author} {\bibfnamefont {V.}~\bibnamefont {Meyer}}, \bibinfo
  {author} {\bibfnamefont {D.}~\bibnamefont {Lucas}}, \bibinfo {author}
  {\bibfnamefont {M.}~\bibnamefont {Barrett}}, \bibinfo {author} {\bibfnamefont
  {J.}~\bibnamefont {Britton}}, \bibinfo {author} {\bibfnamefont {W.~M.}\
  \bibnamefont {Itano}}, \bibinfo {author} {\bibfnamefont {B.}~\bibnamefont
  {Jelenkovi{\'c}}}, \bibinfo {author} {\bibfnamefont {C.}~\bibnamefont
  {Langer}}, \bibinfo {author} {\bibfnamefont {T.}~\bibnamefont {Rosenband}}, \
  and\ \bibinfo {author} {\bibfnamefont {D.~J.}\ \bibnamefont {Wineland}},\
  }\href@noop {} {\bibfield  {journal} {\bibinfo  {journal} {Nature}\ }\textbf
  {\bibinfo {volume} {422}},\ \bibinfo {pages} {412} (\bibinfo {year}
  {2003}{\natexlab{a}})}\BibitemShut {NoStop}%
\bibitem [{\citenamefont {Benhelm}\ \emph {et~al.}(2008)\citenamefont
  {Benhelm}, \citenamefont {Kirchmair}, \citenamefont {Roos},\ and\
  \citenamefont {Blatt}}]{Benhelm-Blatt-low-gate-error}%
  \BibitemOpen
  \bibfield  {author} {\bibinfo {author} {\bibfnamefont {J.}~\bibnamefont
  {Benhelm}}, \bibinfo {author} {\bibfnamefont {G.}~\bibnamefont {Kirchmair}},
  \bibinfo {author} {\bibfnamefont {C.~F.}\ \bibnamefont {Roos}}, \ and\
  \bibinfo {author} {\bibfnamefont {R.}~\bibnamefont {Blatt}},\ }\href@noop {}
  {\bibfield  {journal} {\bibinfo  {journal} {Nature Phys.}\ }\textbf {\bibinfo
  {volume} {4}},\ \bibinfo {pages} {463} (\bibinfo {year} {2008})},\ \Eprint
  {http://arxiv.org/abs/arXiv:0803.2798} {arXiv:0803.2798} \BibitemShut
  {NoStop}%
\bibitem [{\citenamefont {Leibfried}\ \emph
  {et~al.}(2003{\natexlab{b}})\citenamefont {Leibfried}, \citenamefont {Blatt},
  \citenamefont {Monroe},\ and\ \citenamefont
  {Wineland}}]{Leibfried-Wineland-single-ion-quantum-rev}%
  \BibitemOpen
  \bibfield  {author} {\bibinfo {author} {\bibfnamefont {D.}~\bibnamefont
  {Leibfried}}, \bibinfo {author} {\bibfnamefont {R.}~\bibnamefont {Blatt}},
  \bibinfo {author} {\bibfnamefont {C.}~\bibnamefont {Monroe}}, \ and\ \bibinfo
  {author} {\bibfnamefont {D.}~\bibnamefont {Wineland}},\ }\href@noop {}
  {\bibfield  {journal} {\bibinfo  {journal} {Rev. Mod. Phys.}\ }\textbf
  {\bibinfo {volume} {75}},\ \bibinfo {pages} {281} (\bibinfo {year}
  {2003}{\natexlab{b}})}\BibitemShut {NoStop}%
\bibitem [{\citenamefont {Ospelkaus}\ \emph {et~al.}(2011)\citenamefont
  {Ospelkaus}, \citenamefont {Warring}, \citenamefont {Colombe}, \citenamefont
  {Brown}, \citenamefont {Amini}, \citenamefont {Leibfried},\ and\
  \citenamefont {J.Wineland}}]{Ospelkaus-Wineland-microwave-quantum-logic}%
  \BibitemOpen
  \bibfield  {author} {\bibinfo {author} {\bibfnamefont {C.}~\bibnamefont
  {Ospelkaus}}, \bibinfo {author} {\bibfnamefont {U.}~\bibnamefont {Warring}},
  \bibinfo {author} {\bibfnamefont {Y.}~\bibnamefont {Colombe}}, \bibinfo
  {author} {\bibfnamefont {K.~R.}\ \bibnamefont {Brown}}, \bibinfo {author}
  {\bibfnamefont {J.~M.}\ \bibnamefont {Amini}}, \bibinfo {author}
  {\bibfnamefont {D.}~\bibnamefont {Leibfried}}, \ and\ \bibinfo {author}
  {\bibfnamefont {D.}~\bibnamefont {J.Wineland}},\ }\href@noop {} {\bibfield
  {journal} {\bibinfo  {journal} {Nature}\ }\textbf {\bibinfo {volume} {476}},\
  \bibinfo {pages} {181} (\bibinfo {year} {2011})}\BibitemShut {NoStop}%
\bibitem [{\citenamefont {Timoney}\ \emph {et~al.}(2011)\citenamefont
  {Timoney}, \citenamefont {Baumgart}, \citenamefont {Johanning}, \citenamefont
  {Var{\'o}n}, \citenamefont {Plenio}, \citenamefont {Retzker},\ and\
  \citenamefont {Wunderlich}}]{Timoney-Wunderlich-microwave-quantum-logic}%
  \BibitemOpen
  \bibfield  {author} {\bibinfo {author} {\bibfnamefont {N.}~\bibnamefont
  {Timoney}}, \bibinfo {author} {\bibfnamefont {I.}~\bibnamefont {Baumgart}},
  \bibinfo {author} {\bibfnamefont {M.}~\bibnamefont {Johanning}}, \bibinfo
  {author} {\bibfnamefont {A.~F.}\ \bibnamefont {Var{\'o}n}}, \bibinfo {author}
  {\bibfnamefont {M.~B.}\ \bibnamefont {Plenio}}, \bibinfo {author}
  {\bibfnamefont {A.}~\bibnamefont {Retzker}}, \ and\ \bibinfo {author}
  {\bibfnamefont {C.}~\bibnamefont {Wunderlich}},\ }\href@noop {} {\bibfield
  {journal} {\bibinfo  {journal} {Nature}\ }\textbf {\bibinfo {volume} {476}},\
  \bibinfo {pages} {185} (\bibinfo {year} {2011})}\BibitemShut {NoStop}%
\bibitem [{\citenamefont {Chiaverini}\ \emph {et~al.}(2005)\citenamefont
  {Chiaverini}, \citenamefont {Blakestad}, \citenamefont {Britton},
  \citenamefont {Jost}, \citenamefont {Langer}, \citenamefont {Leibfried},
  \citenamefont {Ozeri},\ and\ \citenamefont
  {Wineland}}]{Chiaverini-Wineland-surface-trap-design}%
  \BibitemOpen
  \bibfield  {author} {\bibinfo {author} {\bibfnamefont {J.}~\bibnamefont
  {Chiaverini}}, \bibinfo {author} {\bibfnamefont {R.~B.}\ \bibnamefont
  {Blakestad}}, \bibinfo {author} {\bibfnamefont {J.}~\bibnamefont {Britton}},
  \bibinfo {author} {\bibfnamefont {J.~D.}\ \bibnamefont {Jost}}, \bibinfo
  {author} {\bibfnamefont {C.}~\bibnamefont {Langer}}, \bibinfo {author}
  {\bibfnamefont {D.}~\bibnamefont {Leibfried}}, \bibinfo {author}
  {\bibfnamefont {R.}~\bibnamefont {Ozeri}}, \ and\ \bibinfo {author}
  {\bibfnamefont {D.~J.}\ \bibnamefont {Wineland}},\ }\href@noop {} {\bibfield
  {journal} {\bibinfo  {journal} {Quant. Info. Comp.}\ }\textbf {\bibinfo
  {volume} {5}},\ \bibinfo {pages} {419} (\bibinfo {year} {2005})}\BibitemShut
  {NoStop}%
\bibitem [{\citenamefont {Seidelin}\ \emph {et~al.}(2006)\citenamefont
  {Seidelin}, \citenamefont {Chiaverini}, \citenamefont {Reichle},
  \citenamefont {Bollinger}, \citenamefont {Leibfried}, \citenamefont
  {Britton}, \citenamefont {Wesenberg}, \citenamefont {Blakestad},
  \citenamefont {Epstein}, \citenamefont {Hume}, \citenamefont {Itano},
  \citenamefont {Jost}, \citenamefont {Langer}, \citenamefont {Ozeri},
  \citenamefont {Shiga},\ and\ \citenamefont
  {Wineland}}]{Seidelin-Wineland-surface-trap}%
  \BibitemOpen
  \bibfield  {author} {\bibinfo {author} {\bibfnamefont {S.}~\bibnamefont
  {Seidelin}}, \bibinfo {author} {\bibfnamefont {J.}~\bibnamefont
  {Chiaverini}}, \bibinfo {author} {\bibfnamefont {R.}~\bibnamefont {Reichle}},
  \bibinfo {author} {\bibfnamefont {J.~J.}\ \bibnamefont {Bollinger}}, \bibinfo
  {author} {\bibfnamefont {D.}~\bibnamefont {Leibfried}}, \bibinfo {author}
  {\bibfnamefont {J.}~\bibnamefont {Britton}}, \bibinfo {author} {\bibfnamefont
  {J.~H.}\ \bibnamefont {Wesenberg}}, \bibinfo {author} {\bibfnamefont {R.~B.}\
  \bibnamefont {Blakestad}}, \bibinfo {author} {\bibfnamefont {R.~J.}\
  \bibnamefont {Epstein}}, \bibinfo {author} {\bibfnamefont {D.~B.}\
  \bibnamefont {Hume}}, \bibinfo {author} {\bibfnamefont {W.~M.}\ \bibnamefont
  {Itano}}, \bibinfo {author} {\bibfnamefont {J.~D.}\ \bibnamefont {Jost}},
  \bibinfo {author} {\bibfnamefont {C.}~\bibnamefont {Langer}}, \bibinfo
  {author} {\bibfnamefont {R.}~\bibnamefont {Ozeri}}, \bibinfo {author}
  {\bibfnamefont {N.}~\bibnamefont {Shiga}}, \ and\ \bibinfo {author}
  {\bibfnamefont {D.~J.}\ \bibnamefont {Wineland}},\ }\href {\doibase
  10.1103/PhysRevLett.96.253003} {\bibfield  {journal} {\bibinfo  {journal}
  {Phys. Rev. Lett.}\ }\textbf {\bibinfo {volume} {96}},\ \bibinfo {pages}
  {253003} (\bibinfo {year} {2006})}\BibitemShut {NoStop}%
\bibitem [{\citenamefont {Kim}\ \emph {et~al.}(2011)\citenamefont {Kim},
  \citenamefont {Vlahacos}, \citenamefont {Hoffman}, \citenamefont {Grover},
  \citenamefont {Voigt}, \citenamefont {Cooper}, \citenamefont {Ballard},
  \citenamefont {Palmer}, \citenamefont {Hafezi}, \citenamefont {Taylor},
  \citenamefont {Anderson}, \citenamefont {Dragt}, \citenamefont {Lobb},
  \citenamefont {Orozco}, \citenamefont {Rolston},\ and\ \citenamefont
  {Wellstood}}]{Kim-Wellstood-Rb-hyperfine-SC-resonator}%
  \BibitemOpen
  \bibfield  {author} {\bibinfo {author} {\bibfnamefont {Z.}~\bibnamefont
  {Kim}}, \bibinfo {author} {\bibfnamefont {C.~P.}\ \bibnamefont {Vlahacos}},
  \bibinfo {author} {\bibfnamefont {J.~E.}\ \bibnamefont {Hoffman}}, \bibinfo
  {author} {\bibfnamefont {J.~A.}\ \bibnamefont {Grover}}, \bibinfo {author}
  {\bibfnamefont {K.~D.}\ \bibnamefont {Voigt}}, \bibinfo {author}
  {\bibfnamefont {B.~K.}\ \bibnamefont {Cooper}}, \bibinfo {author}
  {\bibfnamefont {C.~J.}\ \bibnamefont {Ballard}}, \bibinfo {author}
  {\bibfnamefont {B.~S.}\ \bibnamefont {Palmer}}, \bibinfo {author}
  {\bibfnamefont {M.}~\bibnamefont {Hafezi}}, \bibinfo {author} {\bibfnamefont
  {J.~M.}\ \bibnamefont {Taylor}}, \bibinfo {author} {\bibfnamefont {J.~R.}\
  \bibnamefont {Anderson}}, \bibinfo {author} {\bibfnamefont {A.~J.}\
  \bibnamefont {Dragt}}, \bibinfo {author} {\bibfnamefont {C.~J.}\ \bibnamefont
  {Lobb}}, \bibinfo {author} {\bibfnamefont {L.~A.}\ \bibnamefont {Orozco}},
  \bibinfo {author} {\bibfnamefont {S.~L.}\ \bibnamefont {Rolston}}, \ and\
  \bibinfo {author} {\bibfnamefont {F.~C.}\ \bibnamefont {Wellstood}},\ }\href
  {\doibase DOI:10.1063/1.3651466} {\bibfield  {journal} {\bibinfo  {journal}
  {AIP Adv.}\ }\textbf {\bibinfo {volume} {1}},\ \bibinfo {pages} {042107}
  (\bibinfo {year} {2011})}\BibitemShut {NoStop}%
\bibitem [{\citenamefont {Labaziewicz}\ \emph {et~al.}(2008)\citenamefont
  {Labaziewicz}, \citenamefont {Ge}, \citenamefont {Antohi}, \citenamefont
  {Leibrandt}, \citenamefont {Brown},\ and\ \citenamefont
  {Chuang}}]{Labaziewicz-Chuang-cryogenically-suppressed-heating}%
  \BibitemOpen
  \bibfield  {author} {\bibinfo {author} {\bibfnamefont {J.}~\bibnamefont
  {Labaziewicz}}, \bibinfo {author} {\bibfnamefont {Y.}~\bibnamefont {Ge}},
  \bibinfo {author} {\bibfnamefont {P.}~\bibnamefont {Antohi}}, \bibinfo
  {author} {\bibfnamefont {D.}~\bibnamefont {Leibrandt}}, \bibinfo {author}
  {\bibfnamefont {K.~R.}\ \bibnamefont {Brown}}, \ and\ \bibinfo {author}
  {\bibfnamefont {I.~L.}\ \bibnamefont {Chuang}},\ }\href@noop {} {\bibfield
  {journal} {\bibinfo  {journal} {Phys. Rev. Lett.}\ }\textbf {\bibinfo
  {volume} {100}},\ \bibinfo {pages} {013001} (\bibinfo {year}
  {2008})}\BibitemShut {NoStop}%
\bibitem [{\citenamefont {Turchette}\ \emph {et~al.}(2000)\citenamefont
  {Turchette} \emph {et~al.}}]{Turchette-Wineland-ion-heating}%
  \BibitemOpen
  \bibfield  {author} {\bibinfo {author} {\bibfnamefont {Q.~A.}\ \bibnamefont
  {Turchette}} \emph {et~al.},\ }\href@noop {} {\bibfield  {journal} {\bibinfo
  {journal} {Phys. Rev. A}\ }\textbf {\bibinfo {volume} {61}},\ \bibinfo
  {pages} {063418} (\bibinfo {year} {2000})}\BibitemShut {NoStop}%
\bibitem [{\citenamefont {Keller}\ \emph {et~al.}(2004)\citenamefont {Keller},
  \citenamefont {Lange}, \citenamefont {Hayasaka}, \citenamefont {Lange},\ and\
  \citenamefont {Walther}}]{Keller-Walther-ion-cavity-single-photon}%
  \BibitemOpen
  \bibfield  {author} {\bibinfo {author} {\bibfnamefont {M.}~\bibnamefont
  {Keller}}, \bibinfo {author} {\bibfnamefont {B.}~\bibnamefont {Lange}},
  \bibinfo {author} {\bibfnamefont {K.}~\bibnamefont {Hayasaka}}, \bibinfo
  {author} {\bibfnamefont {W.}~\bibnamefont {Lange}}, \ and\ \bibinfo {author}
  {\bibfnamefont {H.}~\bibnamefont {Walther}},\ }\href@noop {} {\bibfield
  {journal} {\bibinfo  {journal} {Nature}\ }\textbf {\bibinfo {volume} {431}},\
  \bibinfo {pages} {1075} (\bibinfo {year} {2004})}\BibitemShut {NoStop}%
\bibitem [{\citenamefont {Olmschenk}\ \emph {et~al.}(2009)\citenamefont
  {Olmschenk}, \citenamefont {Matsukevich}, \citenamefont {Maunz},
  \citenamefont {Hayes}, \citenamefont {Duan},\ and\ \citenamefont
  {Monroe}}]{Olmschenk-Monroe-separated-ion-teleportation}%
  \BibitemOpen
  \bibfield  {author} {\bibinfo {author} {\bibfnamefont {S.}~\bibnamefont
  {Olmschenk}}, \bibinfo {author} {\bibfnamefont {D.~N.}\ \bibnamefont
  {Matsukevich}}, \bibinfo {author} {\bibfnamefont {P.}~\bibnamefont {Maunz}},
  \bibinfo {author} {\bibfnamefont {D.}~\bibnamefont {Hayes}}, \bibinfo
  {author} {\bibfnamefont {L.-M.}\ \bibnamefont {Duan}}, \ and\ \bibinfo
  {author} {\bibfnamefont {C.}~\bibnamefont {Monroe}},\ }\href {\doibase
  10.1126/science.1167209} {\bibfield  {journal} {\bibinfo  {journal}
  {Science}\ }\textbf {\bibinfo {volume} {323}},\ \bibinfo {pages} {486}
  (\bibinfo {year} {2009})}\BibitemShut {NoStop}%
\bibitem [{\citenamefont {Milburn}\ \emph {et~al.}(2000)\citenamefont
  {Milburn}, \citenamefont {Schneider},\ and\ \citenamefont
  {James}}]{Milburn-James-phase-gate}%
  \BibitemOpen
  \bibfield  {author} {\bibinfo {author} {\bibfnamefont {G.~J.}\ \bibnamefont
  {Milburn}}, \bibinfo {author} {\bibfnamefont {S.}~\bibnamefont {Schneider}},
  \ and\ \bibinfo {author} {\bibfnamefont {D.~F.~V.}\ \bibnamefont {James}},\
  }\href@noop {} {\bibfield  {journal} {\bibinfo  {journal} {Fortschr. Phys.}\
  }\textbf {\bibinfo {volume} {48}},\ \bibinfo {pages} {801} (\bibinfo {year}
  {2000})}\BibitemShut {NoStop}%
\bibitem [{\citenamefont {Barenco}\ \emph {et~al.}(1995)\citenamefont
  {Barenco}, \citenamefont {Bennett}, \citenamefont {Cleve}, \citenamefont
  {DiVincenzo}, \citenamefont {Margolus}, \citenamefont {Shor}, \citenamefont
  {Sleator}, \citenamefont {Smolin},\ and\ \citenamefont
  {Weinfurter}}]{Barenco-Weinfurter-universal-gate-set}%
  \BibitemOpen
  \bibfield  {author} {\bibinfo {author} {\bibfnamefont {A.}~\bibnamefont
  {Barenco}}, \bibinfo {author} {\bibfnamefont {C.~H.}\ \bibnamefont
  {Bennett}}, \bibinfo {author} {\bibfnamefont {R.}~\bibnamefont {Cleve}},
  \bibinfo {author} {\bibfnamefont {D.~P.}\ \bibnamefont {DiVincenzo}},
  \bibinfo {author} {\bibfnamefont {N.}~\bibnamefont {Margolus}}, \bibinfo
  {author} {\bibfnamefont {P.}~\bibnamefont {Shor}}, \bibinfo {author}
  {\bibfnamefont {T.}~\bibnamefont {Sleator}}, \bibinfo {author} {\bibfnamefont
  {J.~A.}\ \bibnamefont {Smolin}}, \ and\ \bibinfo {author} {\bibfnamefont
  {H.}~\bibnamefont {Weinfurter}},\ }\href@noop {} {\bibfield  {journal}
  {\bibinfo  {journal} {Phys. Rev. A}\ }\textbf {\bibinfo {volume} {52}},\
  \bibinfo {pages} {3457} (\bibinfo {year} {1995})}\BibitemShut {NoStop}%
\bibitem [{\citenamefont {Munro}\ \emph {et~al.}(2002)\citenamefont {Munro},
  \citenamefont {Nemoto}, \citenamefont {Milburn},\ and\ \citenamefont
  {Braunstein}}]{Munro-Braunstein-schrodinger-cat-heisenberg-limit}%
  \BibitemOpen
  \bibfield  {author} {\bibinfo {author} {\bibfnamefont {W.~J.}\ \bibnamefont
  {Munro}}, \bibinfo {author} {\bibfnamefont {K.}~\bibnamefont {Nemoto}},
  \bibinfo {author} {\bibfnamefont {G.~J.}\ \bibnamefont {Milburn}}, \ and\
  \bibinfo {author} {\bibfnamefont {S.~L.}\ \bibnamefont {Braunstein}},\ }\href
  {\doibase 10.1103/PhysRevA.66.023819} {\bibfield  {journal} {\bibinfo
  {journal} {Phys. Rev. A}\ }\textbf {\bibinfo {volume} {66}},\ \bibinfo
  {pages} {023819} (\bibinfo {year} {2002})}\BibitemShut {NoStop}%
\bibitem [{\citenamefont {M{\o}lmer}\ and\ \citenamefont
  {S{\o}rensen}(1999)}]{Molmer-Sorensen-multiparticle-entangling-gate}%
  \BibitemOpen
  \bibfield  {author} {\bibinfo {author} {\bibfnamefont {K.}~\bibnamefont
  {M{\o}lmer}}\ and\ \bibinfo {author} {\bibfnamefont {A.}~\bibnamefont
  {S{\o}rensen}},\ }\href {\doibase 10.1103/PhysRevLett.82.1835} {\bibfield
  {journal} {\bibinfo  {journal} {Phys. Rev. Lett.}\ }\textbf {\bibinfo
  {volume} {82}},\ \bibinfo {pages} {1835} (\bibinfo {year}
  {1999})}\BibitemShut {NoStop}%
\bibitem [{\citenamefont {S{\o}rensen}\ and\ \citenamefont
  {M{\o}lmer}(2000)}]{Sorensen-Molmer-bichromatic-gate-PRA}%
  \BibitemOpen
  \bibfield  {author} {\bibinfo {author} {\bibfnamefont {A.}~\bibnamefont
  {S{\o}rensen}}\ and\ \bibinfo {author} {\bibfnamefont {K.}~\bibnamefont
  {M{\o}lmer}},\ }\href {\doibase 10.1103/PhysRevA.62.022311} {\bibfield
  {journal} {\bibinfo  {journal} {Phys. Rev. A}\ }\textbf {\bibinfo {volume}
  {62}},\ \bibinfo {pages} {022311} (\bibinfo {year} {2000})}\BibitemShut
  {NoStop}%
\end{thebibliography}
\end{document}


\title{Quantum interface between an electrical circuit and a single atom \\ Supplemental material: Details of technical calculations}
\author{D. Kielpinski}
\affiliation{Centre for Quantum Dynamics, Griffith University, Nathan, QLD 4111, Australia}
\author{D. Kafri}
\affiliation{Joint Quantum Institute/NIST, College Park, MD, USA}
\author{M. J. Woolley}
\affiliation{Centre for Engineered Quantum Systems, School of Mathematics and Physics, The University of Queensland, St Lucia, QLD 4072, Australia}
\author{G. J. Milburn}
\affiliation{Centre for Engineered Quantum Systems, School of Mathematics and Physics, The University of Queensland, St Lucia, Australia 4072}
\author{J. M. Taylor}
\affiliation{Joint Quantum Institute/NIST, College Park, MD, USA}

\maketitle

\section{Technical details of ion-circuit coupling device}

\subsection{Electrical properties}

The inductor attached to the island electrodes consists of a 30 turn coil of superconducting wire with diameter $d = 1$ mm and length $\ell = 650 \:\mu$m, providing an inductance of 440 nH. The total static circuit capacitance of 46 fF arises from the parasitic capacitances of the inductor and the static capacitance of the BAW. The chief parasitic effect on the LC circuit is the capacitive coupling of the inductor to nearby metallic surfaces, approximated here as a long cylindrical shield of diameter $D = 5$ mm surrounding the coil. The shield capacitance is then given by $2 \pi \epsilon_0 \ell / \ln (D/d) = 16$ fF. The self-capacitance of the coil can be calculated in the quasistatic approximation \cite{Massarini-Kazimierczuk-inductor-self-capacitance}. Assuming a wire diameter of 50 $\mu$m and gaps of 15 $\mu$m around each winding, the self-capacitance is found to be negligible, on the order of a few fF. The capacitance between the trap island electrodes is also negligible, as it is much smaller than the few fF self-capacitance of either island \cite{Smythe-static-dynamic-electricity-BOOK}. The high impedance of the LC circuit (2.7 k$\Omega$) increases the single-photon electric field to $> 7$ times that of the coplanar waveguides generally used in circuit QED \cite{Wallraff-Schoelkopf-strong-coupling-CQED} without loss of resonator $Q$, which can remain as high as $10^5$ \cite{Teufel-Simmonds-strong-coupling-SC-electromechanics}. \\

The island electrodes are taken to be square with side $R = 50 \: \mu$m and a gap of $s = 10 \: \mu$m normal to the long axis of the trap. The electric field due to the island electrodes is evaluated by approximating the islands as long, wide conducting strips extending along $y$. This approximation is justified since the islands are reasonably large relative to the gap between them and relative to the height of the ion. The inter-island capacitance can be evaluated as a few pF and an analytic series expansion gives the geometric factor $\zeta = 0.25$ \cite{Simons-Arora-CPW-field-components}. We tested the validity of our approximation by varying the ion height, electrode width, and gap width over a factor of two. The value of $\zeta$ varied by no more than $\pm 20$\% as long as the ion height was less than the electrode width, confirming that our result is robust. \\

\subsection{BAW properties}

A silicon bulk-acoustic-wave resonator (BAW) is mounted near the inductor with superconducting metallic electrodes that are separated from the inductor terminals by a submicron gap. Acousto-electric structures of this type are now widely used in microwave circuitry as resonators and filter components \cite{Weigel-Ruppel-microwave-acoustic-device-rev} and exhibit sharp mechanical resonances with $Q \sim 10^4$ at frequencies near 1 GHz \cite{Pourkamali-Ayazi-silicon-MEMS-resonator-design}. Acoustic waves are excited in the BAW by applying a classical voltage at frequency $\nu_B \approx \omega_\mathrm{LC} - \omega_i$ to a secondary electrode. The resulting modulation of the gap distance between the BAW electrodes and the superconducting inductor terminals results in a modulation of the LC capacitance. The BAW excitation electrode is placed far from the capacitance-modulation electrodes, so there is negligible cross-coupling between the classical BAW drive and the LC circuit. \\

A BAW structure of this type can easily modulate the LC resonator frequency with the desired modulation depth of $\eta = 0.3$. For a typical BAW dimension of 30 $\mu$m, the static capacitance is $C_\mathrm{B0} = 30$ fF, so the BAW gap is the chief contributor of capacitance in the circuit. Current technology allows the static gap distance between BAW and substrate to be made as small as $\zeta_0 \sim 100$ nm \cite{Pourkamali-Ayazi-silicon-MEMS-100nm-gap}, much less than the vibrational amplitude allowed by material strain \cite{Kaajakari-Seppa-silicon-NEMS-nonlinearity}. Modulating the BAW gap distance $\zeta(t)$ can therefore impose a large modulation on the total capacitance. The use of superconducting electrodes for the BAW device ensures minimal electrical dissipation in the circuit even under driving. \\

\section{Derivation of parametric LC-motion coupling}

We start by quantizing the classical time-dependent LC Hamiltonian $H_\mathrm{LC,cl} = \Phi^2/(2 L) + (1 - \eta \sin \nu t) Q^2/(2 C_0)$. Here $\Phi$ denotes the flux in the LC. Following the method of Brown, we make a canonical transformation defined using the solutions to the classical LC Hamiltonian to obtain the quantum Hamiltonian in terms of quasienergy states. While we assume $\eta \ll 1$ below for clarity, our derivation still holds for arbitrary $\eta < 1$ if the full Mathieu-function classical solutions are used. The classical solutions are given at ${\mathcal O}(\eta)$ by
\beq
q_\pm(t) = e^{\pm i \omega t} - \sfrac{1}{6} \: \eta (e^{\pm 2 i \omega t} + 3 e^{\mp i (\nu - \omega) t})
\eeq
where we approximate $\nu + \omega \approx 2 \omega$ since $\nu, \omega \gg |\nu - \omega|$. The quantum LC Hamiltonian is then found to be
\beq
H_\mathrm{LC}/\hbar = (1 + \sfrac{2}{3} \: \eta \sin \nu t) \left[ A^\dagger A + \sfrac{1}{2} \right]
\eeq
where $A(t)$ is the annihilation operator for quasienergy states with commutation relation $[A, A^\dagger] = 1 + {\mathcal O}(\eta^2)$. The free-field solution of $H_\mathrm{LC}$ is $A(t) = e^{-i \omega t} (1 - \sfrac{2}{3} \: i \eta \sin \nu t) A(0)$. The original charge operator can now be written as
\beq
Q = (1 - \sfrac{1}{3} \: \eta \sin \nu t) \sqrt{\frac{\hbar}{2 Z}} \left[ A(t) + A^\dagger(t) \right]
\eeq
Neglecting zero-point energy terms, the total Hamiltonian of the ion-LC system becomes
\ba
H/\hbar &=& \omega_\mathrm{LC} (1 + \sfrac{2}{3} \: \eta \sin \nu t) A^\dagger A + \omega_i B^\dagger B \nonumber \\
&& \quad + g_0 (1 + \sfrac{2}{3} \: \eta \sin \nu t) (A + A^\dagger) (B + B^\dagger)
\label{ham}
\ea
where the coupling constant $g_0 = e \zeta z_0 q_0 / (h C_0)$ and $B$ is the annihilation operator of the ion motion. For the parameters given above, $g_0 = 2 \pi \times 200$ kHz. \\

To remove the free-field terms in Eq. (\ref{ham}), we perform a canonical transformation implemented by the unitary operator
\beq
\label{transform}U = \exp \left[ i \omega_\mathrm{LC} \left( t - \frac{2\eta}{3\nu} \cos \nu t \right) A^\dagger A + i \omega_i t B^\dagger B \right] \eeq To lowest order in $\eta$, the Hamiltonian becomes
\ba
H_\mathrm{int}/\hbar &=& g_0 [ e^{-i \omega_\mathrm{LC} t} \kappa(t) a + \mbox{h.c.} ] (b e^{-i \omega_i t} + \mbox{h.c.}) \label{intframe} \\
\kappa(t) &\equiv& 1 + (2\eta/3) (\sin \nu t - i (\omega_i/\nu) \cos \nu t)
\ea
The transformed LC and motional operators $a, b$ are now slowly varying with respect to $\omega_\mathrm{LC}, \omega_i$, and $\nu$. We define the detuning $\Delta \equiv \nu - (\omega_\mathrm{LC} - \omega_i)$ and make a rotating-wave approximation to find the parametric resonance Hamiltonian
\beq
H_\mathrm{int}/\hbar = \frac{2 i g_0 \eta}{3} e^{-i \Delta t} a b^\dagger + \mbox{h.c.} \label{intframe2}
\eeq
where the counterrotating terms give rise to corrections of order $\eta g/\nu$ and we have neglected contributions of order $\Delta/\omega_i$ in the prefactor. Extending the calculation to $\mathcal{O}(\eta^2)$ adds terms with frequencies $\omega_\mathrm{LC} - \omega_i \pm 2 \nu$ to the Hamiltonian. These terms are nonresonant, so Eq. (\ref{intframe}) remains valid even when $\eta$ is a significant fraction of 1. \\